\documentclass[aps,pre,reprint,floatfix,superscriptaddress]{revtex4-1}
\pdfoutput=1
\usepackage{amsmath}
\usepackage{amssymb}
\usepackage{graphicx}
\usepackage[colorlinks=true,linkcolor=blue,citecolor=blue]{hyperref}

\newcommand{\mshear}{s_i}
\newcommand{\kperp}{k_\perp}
\newcommand{\qo}{\tilde{q}_0}
\newcommand{\sym}[1]{\hat{#1}}
\newcommand{\solp}{S_\text{p}}
\newcommand{\solf}{S_\text{F}}
\newcommand{\sold}{S_\text{d}}
\newcommand{\kpar}{k_\parallel}
\newcommand{\asym}[1]{\check{#1}}
\newcommand{\Psib}{\Psi_\text{b}}
\newcommand{\miota}{\tilde{\iota}}
\newcommand{\qb}{\tilde{q}_\text{b}}
\newcommand{\helena}{\texttt{HELENA}}
\newcommand{\ansatz}{\textit{ansatz}}
\newcommand{\rhopol}{\rho_\text{pol}}

\begin{document}

\title{
Local up-down asymmetrically shaped equilibrium model for tokamak plasmas}

\date{\today}

\author{Paulo Rodrigues}
\affiliation{
Instituto de Plasmas e Fus\~{a}o Nuclear, Instituto Superior T\'{e}cnico,
Universidade de Lisboa, 1049-001 Lisboa, Portugal.}

\author{Andr\'{e} Coroado}
\affiliation{
Instituto de Plasmas e Fus\~{a}o Nuclear, Instituto Superior T\'{e}cnico,
Universidade de Lisboa, 1049-001 Lisboa, Portugal.}
\affiliation{
\'{E}cole Polytechnique F\'{e}d\'{e}rale de Lausanne (EPFL), Swiss Plasma Center
(SPC), CH-1015 Lausanne, Switzerland.}

\begin{abstract}
A local magnetic equilibrium model is presented, with finite inverse aspect
ratio and up-down asymmetrically shaped cross section, that depends on eight
free parameters. In contrast with other local equilibria, which provide simple
magnetic-surface parametrisations at the cost of complex poloidal-field flux
descriptions, the proposed model is intentionally built to afford analytically
tractable magnetic-field components. Therefore, it is particularly suitable for
analytical assessments of equilibrium-shaping effects on a variety of
tokamak-plasma phenomena.
\end{abstract}

\maketitle

\section{Introduction}

Although magnetic equilibria give support to virtually every phenomena in
tokamak plasmas, accurate numerical solutions of the Grad-Shafranov (GS)
equation are not always the best tool to understand or gain insight into such
complex processes. Simplified descriptions are often preferable, either to
achieve analytically tractable expressions or to perform parameter scans without
the need to recompute a numerical equilibrium at every step. With this aim in
mind, local equilibrium models have been developed over the past years and have
seen a wide range of applications: Among others, these include analytical
studies on stability (e.g., ballooning
modes~\cite{connor.1978,greene.1981,bishop.1986}, Alfv\'{e}n
eigenmodes~\cite{fu.1989,candy.1996,berk.2001}, zonal
flows~\cite{rosenbluth.1998,hinton.1999}) and charged-particle
orbits~\cite{wong.1995,roach.1995,brizard.2011}, as well as large-scale
numerical simulations carried out with gyrokinetic
codes~\cite{dorland.2000,jenko.2000,candy.2003,peeters.2009} to understand
microturbulence and its associated transport of heat, momentum, and particles.
In such large-scale simulations, simple magnetic-field descriptions within a
thin flux-tube domain around a given field line are crucial to reduce the
computational effort~\cite{beer.1995}. Besides axisymmetric configurations,
local equilibrium models have also been developed for the more complex,
three-dimensional stellarator geometry~\cite{hegna.2000}.

Most often, local equilibrium models result from an expansion of the
poloidal-field flux per unit angle $\Psi$ in powers of some radial coordinate
around a magnetic surface of prescribed shape, using the GS equation
\begin{equation}
- R \, \nabla \cdot \bigl( R^{-1} \nabla \Psi \bigr) = \mu_0 R^2 p' + F F'
\label{eq:gs.equation}
\end{equation}
and the axisymmetric magnetic-field definition 
\begin{equation}
\mathbf{B} = \nabla \phi \times \nabla \Psi - F \nabla \phi
\label{eq:axisymmetric.field}
\end{equation}
(with $R$ the distance to the torus axis and $\phi$ the toroidal angle) to
relate the first two series coefficients with the poloidal field and the
derivatives of the pressure $p(\Psi)$ and of the diamagnetic function $F(\Psi)$.
In turn, magnetic-surface descriptions range from shifted circles in the
$s-\alpha$ model~\cite{connor.1978} to more sophisticated shapes of the
type~\cite{miller.1998}
\begin{equation}
\begin{gathered}
R(\rho,\vartheta) = R_0 + \Delta(\rho) + \rho \cos \bigl[ \vartheta +
\sin^{-1} \delta(\rho) \sin \vartheta \bigr], \\
Z(\rho,\vartheta) = \kappa(\rho) \rho \sin \vartheta,
\end{gathered}
\label{eq:miller.parametrisation}
\end{equation}
written in terms of shaping parameters like the Shafranov shift $\Delta$, the
elongation $\kappa$ and the triangularity $\delta$, which are constant over each
magnetic surface labeled by $\rho$. Here, $R_0$ is the magnetic axis position on
the midplane and $Z$ the height above it. The coordinates $(\rho, \vartheta)$
are not orthogonal and the metric-tensor components $g^{\rho \rho} = | \nabla
\rho |^2$, $g^{\rho \vartheta} = \nabla \rho \cdot \nabla \vartheta$, and
$g^{\vartheta \vartheta} = | \nabla \vartheta |^2$, although computable
from~\eqref{eq:miller.parametrisation}, yield intricate
expressions~\cite{zhou.2011} that turn analytical work into a very difficult
task.

In many practical applications, however, details about the magnetic-surfaces'
shape are not as important as it is to obtain simple magnetic-field components
from definition~\eqref{eq:axisymmetric.field}, along with a simple geometry and
metric tensor. To meet these needs, a local equilibrium model is developed in
section~\ref{sec:localeq} that builds upon an analytical form for the poloidal
flux with locally adjustable parameters, instead of a predefined
magnetic-surface shape. Its geometric properties are related with other local
models in section~\ref{sec:geometry} and explicit expressions for the
magnetic-field components are provided. In section~\ref{sec:accuracy}, the
accuracy of the proposed model is tested against a numerical equilibrium, while
its suitability to analytical manipulation is illustrated with a couple of
examples in section~\ref{sec:applications}.

\section{Local equilibrium model}
\label{sec:localeq}

As a first step, magnetic-surface induced coordinates are replaced by the
right-handed set $(r, \theta, \phi)$ defined as
\begin{equation}
R(r, \theta) = R_0 \bigl( 1 + \varepsilon r \cos \theta), \quad
Z(r, \theta) = a r \sin \theta,
\label{eq:lab.coordinates}
\end{equation}
where $r$ is the distance to the magnetic axis normalized to the torus minor
radius $a$ and $\varepsilon = a/R_0$ is the inverse aspect ratio. The metric
tensor is diagonal and its nonzero components are
\begin{equation}
g_{rr} = a^2, \quad g_{\theta\theta} = a^2 r^2, \quad g_{\phi\phi} = R^2,
\label{eq:metric.tensor}
\end{equation}
with $\sqrt{g} = a^2 r R$ the Jacobian. Next, the focus is shifted from a
detailed surface description, as in~\eqref{eq:miller.parametrisation}, to a
suitable parametrisation of the flux $\Psi$. To this end, a global solution of
the GS equation, analytical and depending on a few parameters, is used to
generate a family of local solutions, each one with its parameters locally
adjusted in order to approximate the equilibrium being modelled near a given
magnetic surface. Henceforth, $A_\alpha$ and $A^\alpha$ denote, respectively,
the covariant and the contravariant components of some vector $\mathbf{A}$.

The Solovev model~\cite{solovev.1968} provides the simplest family of analytical
global equilibria, with $p(\Psi)$ and $F^2(\Psi)$ linear in $\Psi$. Two
adimensional constants can be defined as
\begin{equation}
\solp = \mu_0 R_0^4 \Psib^{-1} \, p', \quad \solf = R_0^2 \Psib^{-1} \, F F',
\label{eq:solovev.constants}
\end{equation}
where $\Psib$ is the poloidal flux at the plasma boundary. The covariant
toroidal current density
\begin{equation}
J_\phi(R,\Psi) = - R^2 p' - \mu_0^{-1} FF'
\label{eq:toroidal.current}
\end{equation}
becomes independent of the poloidal flux, that is
\begin{equation}
J_\phi(R) = - \mu_0^{-1} R_0^{-2} \Psib
  \Bigl[ \bigl( R/R_0 \bigr)^2 \solp + \solf \Bigr],
\label{eq:solovev.current}
\end{equation}
and the GS equation can be written as
\begin{equation}
x \partial_x \bigl( x^{-1} \partial_x \psi  \bigr)
    + \partial^2_{yy} \psi = - \bigl( \solf + \solp x^2 \bigr),
\label{eq:solovev.equation}
\end{equation}
where $x = R/R_0$, $y = Z/R_0$ and $\psi =
\Psi/\Psib$~\cite{solovev.1968,cerfon.2010}. The latter can be split as the
sum~\cite{cerfon.2010}
\begin{equation}
\psi(x,y) = - \tfrac{1}{8} \solp x^4 - \tfrac{1}{2} \solf x^2 \ln x
    + \psi_\text{h}(x,y),
\label{eq:solovev.solution.xy}
\end{equation}
with $\psi_\text{h}(x,y)$ an arbitrary linear combination of homogeneous
solutions of equation~\eqref{eq:solovev.equation}. Although $\psi_\text{h}$ can
be expressed as an infinite series involving $\ln x$ and powers of $x$ and
$y$~\cite{zheng.1996}, it is sufficient to keep only a finite number of terms in
order to describe the geometry of tokamak plasmas in a wide range of
conditions~\cite{cerfon.2010}.

Albeit analytically tractable, Solovev equilibria cannot describe most features
of current-density distributions in tokamak experiments. True for global
equilibria, with $\solp$ and $\solf$ strictly constant over the cross section, a
local approach avoids this limitation: within a small region of size $\bigl|
\Delta\psi \bigr|$ around the magnetic surface $\psi_i$ such that
\begin{equation}
\bigl| \Delta\psi \:
    J_\phi^{-1}(R,\psi_i) \: \partial_\psi J_\phi(R,\psi_i) \bigr| \ll 1,
\label{eq:current.condition}
\end{equation}
the relation~\eqref{eq:solovev.current} with constant values $\solp(\psi_i)$ and
$\solf(\psi_i)$ approximates equation~\eqref{eq:toroidal.current} and the
solution~\eqref{eq:solovev.solution.xy} is thus locally valid. It is worth
noticing that $|\Delta\Psi \, p''| \ll |p'|$ and $|\Delta\Psi \, (F^2)''| \ll
|(F^2)'|$, although sufficient to ensure the more general
condition~\eqref{eq:current.condition}, are not actually necessary.

The most general form for $\psi_\text{h}$ that enables one to keep terms up to
$\varepsilon^4 r^4 \ll 1$ is the finite series
\begin{equation}
\psi_\text{h} = \sym{c}_0 + \sum_{i=1}^4
    \sym{c}_i \sym{\psi}_\text{h}^i + \asym{c}_i \asym{\psi}_\text{h}^i,
\label{eq:homogeneous.sum}
\end{equation}
where the symmetric homogeneous harmonics are~\cite{zheng.1996,cerfon.2010}
\begin{equation}
\begin{gathered}
\sym{\psi}_\text{h}^1 = x^2, \quad
\sym{\psi}_\text{h}^2 = y^2 - x^2 \ln x, \quad
\sym{\psi}_\text{h}^3 = x^4 - 4 x^2 y^2, \\
\sym{\psi}_\text{h}^4 = 2 y^4 - 9 y^2 x^2 +
    \bigl( 3 x^4 - 12 x^2 y^2 \bigr) \ln x,
\end{gathered}
\label{eq:homogeneous.symmetric}
\end{equation}
and the asymmetric ones are
\begin{equation}
\begin{gathered}
\asym{\psi}_\text{h}^1 = y, \quad
\asym{\psi}_\text{h}^2 = x^2 y, \quad \\
\asym{\psi}_\text{h}^3 = y^3 - 3 x^2 y \ln x, \quad
\asym{\psi}_\text{h}^4 = 3 x^4 y - 4 x^2 y^3.
\end{gathered}
\label{eq:homogeneous.asymmetric}
\end{equation}
As $\solp$ and $\solf$, the coefficients $\sym{c}_i$ and $\asym{c}_i$ must also
change smoothly, accounting for shaping currents outside $\Delta\psi$.

After converting from $(x, y)$ to $(r, \theta)$ via
transformation~\eqref{eq:lab.coordinates} and eliminating $\sym{c}_0$,
$\sym{c}_1$, and $\asym{c}_1$ with the on-axis conditions $\psi
= \nabla \psi = 0$, the solution~\eqref{eq:solovev.solution.xy} becomes
\begin{equation}
\psi(r, \theta) = S_0 r^2 \Bigl[ \Theta_0(\theta) +
    \varepsilon r \Theta_1(\theta) + \varepsilon^2 r^2 \Theta_2(\theta) \Bigr],
\label{eq:solovev.solution.rtheta}
\end{equation}
where $S_0 = - \tfrac{1}{4} \varepsilon^2 \bigl( \solp + \solf \bigr)$ [and thus
$S_0 \sim \tfrac{1}{2} \qb / \qo$ if the cylindrical limits $\qb = a^2 B_0 \big/
\Psib$ and $\qo = 2 B_0 \big/ \mu_0 J_\phi(R_0)$ of the safety factor are
defined] and
\begin{equation}
\begin{aligned}
\Theta_0(\theta) &= 1 + \sym{\kappa} \cos 2 \theta
    + \asym{\kappa} \sin 2 \theta, \\
\Theta_1(\theta) &= \sym{\Delta} \cos \theta +
    \tfrac{1}{4} \asym{\kappa} \sin \theta +
        \sym{\eta} \cos 3 \theta + \asym{\eta} \sin 3 \theta, \\
\Theta_2(\theta) &=
    \tfrac{1}{32}  \bigl( 8 \sym{\Delta} - 3 \sym{\kappa} - 3 \bigr) +
    \tfrac{1}{8} \bigl( 2 \sym{\eta} +
        2 \sym{\Delta} - \sym{\kappa} - 1 \bigr) \cos 2 \theta \\
    & \quad + \tfrac{1}{16} \bigl( 4 \asym{\eta} -
        \asym{\kappa} \bigr) \sin 2 \theta +
            \sym{\chi} \cos 4 \theta + \asym{\chi} \sin 4 \theta.
\end{aligned}
\label{eq:capital.thetas}
\end{equation}
The geometric coefficients $\sym{\kappa}$, $\asym{\kappa}$, $\sym{\Delta}$,
$\sym{\eta}$, $\asym{\eta}$, $\sym{\chi}$, and $\asym{\chi}$ are related with
$\solp$, $\solf$, and the remaining six constants in the
sum~\eqref{eq:homogeneous.sum} by the linear and invertible transformations
\begin{equation}
\begin{gathered}
\begin{bmatrix}
\solp \\ \solf \\ \sym{c}_2 \\ \sym{c}_3 \\ \sym{c}_4
\end{bmatrix} = \frac{S_0}{\varepsilon^{2}}
\renewcommand*{\arraystretch}{1.2}
\begin{bmatrix}
1 & 1 & -4 & 0 & 0 \\
-5 & -1 & 4 & 0 & 0 \\
\frac{29}{32} & -\frac{35}{32} & -\frac{1}{2} &
    \frac{9}{4} & -3 \\
-\frac{15}{256} & -\frac{15}{256} &
    -\frac{1}{8} & \frac{27}{32} & -\frac{15}{8} \\
\frac{1}{64} & \frac{1}{64} & 0 & -\frac{1}{8} & \frac{1}{2}
\end{bmatrix}
\begin{bmatrix}
1 \\ \sym{\kappa} \\ \sym{\Delta} \\ \sym{\eta} \\ \sym{\chi}
\end{bmatrix} \\
\begin{bmatrix}
\asym{c}_2 \\ \asym{c}_3 \\ \asym{c}_4
\end{bmatrix} = \frac{S_0}{\varepsilon^{2}}
\renewcommand*{\arraystretch}{1.2}
\begin{bmatrix}
\frac{11}{8} & -\frac{3}{2} & 0 \\
\frac{5}{16} & -\frac{5}{4} & 2 \\
\frac{1}{64} & -\frac{1}{16} & \frac{1}{2}
\end{bmatrix}
\begin{bmatrix}
\asym{\kappa} \\ \asym{\eta} \\ \asym{\chi}
\end{bmatrix}.
\end{gathered}
\label{eq:transformation}
\end{equation}
Note that the flux~\eqref{eq:solovev.solution.rtheta} is a particular case of a
general non-local GS \ansatz~\cite{rodrigues.2004,rodrigues.2009}, whose
relation with Solovev equilibria near the axis is already well
established~\cite{rodrigues.2014}.

\section{Geometry and field components}
\label{sec:geometry}

Intuition about the geometric coefficients is found by inverting
equation~\eqref{eq:solovev.solution.rtheta} to get $r(\theta)$ for constant
$\psi$. Letting $r(\theta) = r_0(\theta) + \varepsilon r_1(\theta) +
\varepsilon^2 r_2(\theta) + \cdots$ and collecting the same powers of
$\varepsilon$ after substitution in the flux
distribution~\eqref{eq:solovev.solution.rtheta}, returns an equation for each
contribution $r_i(\theta)$ and, at length, the magnetic-surface parametrisation
\begin{multline}
r(\theta) = \tilde{s} \Biggl( \frac{1}{\Theta_0^{1/2}}
    - \frac{\Theta_1}{2 \Theta_0^2} \, \varepsilon \tilde{s} \\
    + \frac{5 \Theta_1^2 - 4 \Theta_0 \Theta_2}{8
    \Theta_0^{7/2}} \, \varepsilon^2 \tilde{s}^2
    - \Theta_1 \frac{2 \Theta_1^2 - 3 \Theta_0 \Theta_2}{2 \Theta_0^5}
    \, \varepsilon^3 \tilde{s}^3 \\
    + \frac{7}{128} \frac{33 \Theta_1^4 - 72 \Theta_0 \Theta_1^2 \Theta_2 + 16 \Theta_0^2 \Theta_2^2}{\Theta_0^{13/2}}
        \, \varepsilon^4 \tilde{s}^4 + \cdots \Biggr),
\label{eq:magnetic.surface}
\end{multline}
which is accurate to terms of order $\varepsilon^4 \tilde{s}^4$ with $\tilde{s}
= \sqrt{\psi/S_0}$.  The angle $\theta_\text{high}$ of a symmetric surface
highest point [corresponding to $\vartheta = \pi/2$
in~\eqref{eq:miller.parametrisation}] is, at leading order in $\varepsilon$,
\begin{equation}
\cos \theta_\text{high} = -
    \frac{\varepsilon \tilde{s}}{2 \sqrt{1-\sym{\kappa}}}
        \frac{\sym{\Delta} - 3 \sym{\eta}}{1 + \sym{\kappa}} + \cdots.
\label{eq:cos.theta.high}
\end{equation}
Thus, one finds the conventional definitions of $\rho$, $\Delta$, $\kappa$, and
$\delta$~\cite{miller.1998,cerfon.2010} to yield the leading order
approximations
\begin{equation}
\begin{gathered}
\frac{\rho}{a} \approx \frac{\tilde{s}}{\sqrt{1 + \sym{\kappa}}}, \quad
\kappa \approx \sqrt{\frac{1 + \sym{\kappa}}{1 - \sym{\kappa}}}, \\
\frac{\Delta}{a} \approx - \frac{\varepsilon \tilde{s}^2}{2}
    \frac{\sym{\Delta} + \sym{\eta}}{\bigl( 1 + \sym{\kappa} \bigr)^2},\quad
\delta \approx \frac{\rho}{R_0} \frac{
        \sym{\kappa} \bigl( \sym{\Delta} - \sym{\eta} \bigr) - 2 \sym{\eta}}{
        1 - \sym{\kappa}^2}.
\end{gathered}
\label{eq:geometric.equivalences}
\end{equation}
The coefficient $\sym{\chi}$, absent from the relations above, relates with the
surface's quadrangularity, which is not described by
parametrisation~\eqref{eq:miller.parametrisation}. In turn, $\asym{\kappa}$ is
connected with the surface's tilt away from the
vertical~\cite{ball.2014,rodrigues.2014}, whereas $\asym{\eta}$ and
$\asym{\chi}$ provide higher-order asymmetric corrections.

Equation~\eqref{eq:axisymmetric.field} sets the
magnetic-field components on the poloidal plane. If the geometric coefficients depend on the
surface label $\psi$, condition~\eqref{eq:solovev.solution.rtheta} becomes
implicit and $\nabla \psi$ follows from the implicit function theorem: defining
\begin{equation}
H = r^2 \Bigl[ \bigl(S_0 \Theta_0 \bigr)'
  + \varepsilon r \bigl(S_0 \Theta_1 \bigr)'
    + \varepsilon^2 r^2 \bigl(S_0 \Theta_2 \bigr)' \Bigr]
\label{eq:implicit.derivative}
\end{equation}
and $\Theta_i' = \partial_\psi\Theta_i$, the poloidal-field components are thus
\begin{gather}
B^r(r, \theta) = - r \frac{B_0}{R} \frac{S_0}{\qb} \frac{
  \dot{\Theta}_0 +
        \varepsilon r \: \dot{\Theta}_1 +
            \varepsilon^2 r^2 \: \dot{\Theta}_2}{1 - H},
\label{eq:contravariant.Br}\\
B^\theta(r, \theta)  = \frac{B_0}{R} \frac{S_0}{\qb} \frac{2 \Theta_0 +
        3 \varepsilon r \: \Theta_1 + 4 \varepsilon^2 r^2 \: \Theta_2}{1 - H},
\label{eq:contravariant.Btheta}
\end{gather}
with $\dot{\Theta}_i = \partial_\theta\Theta_i$, if $1 - H$ does not vanish. The
linear diamagnetic-function model near each magnetic surface is $F^2 = B_0^2
R_0^2 \bigl( 1 + \varepsilon^2 \sold \psi \bigr)$, with $\sold$ a new local
coefficient, whence the toroidal field
\begin{equation}
B_\phi(r, \theta) =
    B_0 R_0 \sqrt{1 + \varepsilon^2 \sold \psi(r, \theta)}.
\label{eq:covariant.Btor}
\end{equation}

Relations~\eqref{eq:contravariant.Br} to~\eqref{eq:covariant.Btor} involve
linear combinations of products between $r$ powers and trigonometric functions
of $\theta$. On the contrary, equation~(37) in
reference~\onlinecite{miller.1998} shows combinations of the type
$\sin(\vartheta + \sin^{-1} \delta \sin \vartheta)$, which are much harder to
work with analytically. Assuming surface descriptions simpler than
parametrisation~\eqref{eq:miller.parametrisation} avoids this
limitation~\cite{zhou.2011,yu.2012}, but one must, in any case, change from $(R,
Z, \phi)$ to surface-induced coordinates $(\rho, \vartheta, \phi)$. This
requires a non-trivial, non-diagonal metric tensor, more complex than the one in
definition~\eqref{eq:metric.tensor}. Moreover, some parameters in
equation~\eqref{eq:miller.parametrisation} cannot be arbitrarily set, because a
given shape does not necessarily correspond to a magnetic surface of a valid
equilibrium~\cite{yu.2012}. In contrast, any choice of coefficients in
equation~\eqref{eq:solovev.solution.rtheta} yields, via
transformation~\eqref{eq:transformation}, a set of constants in
the \ansatz~\eqref{eq:solovev.solution.xy} which is always a solution of
equation~\eqref{eq:solovev.equation} up to terms of order $\varepsilon^4 r^4$.
On the other hand, the surface description in
equation~\eqref{eq:magnetic.surface} is more complex than
parametrisation~\eqref{eq:miller.parametrisation}, but the benefits of a simpler
magnetic field for analytical work are often more important than the conciseness
of the surface's shape.

\begin{figure}
\includegraphics[scale=1.0]{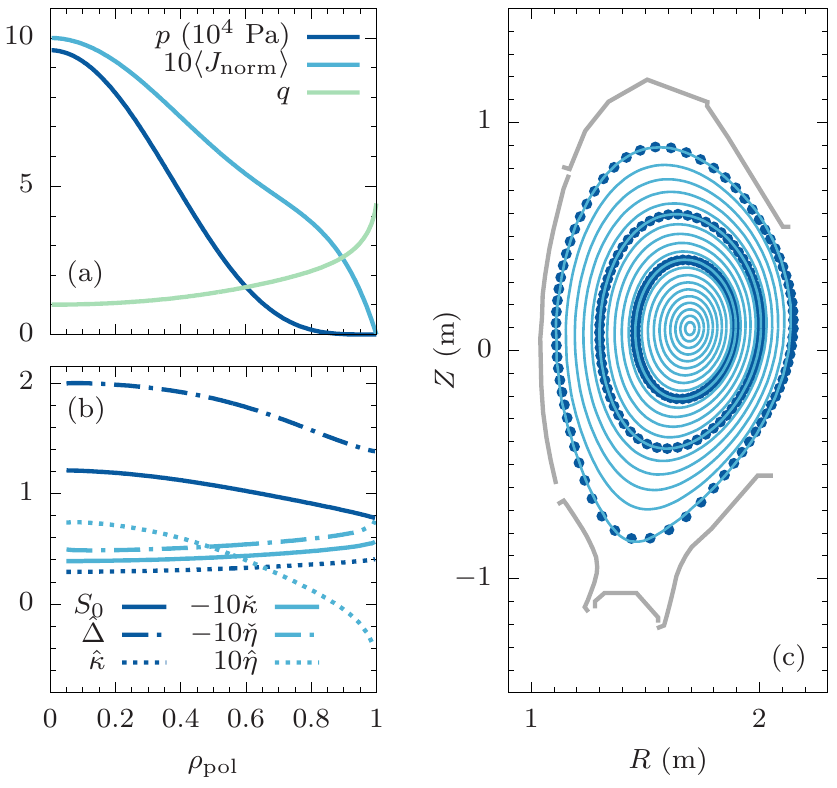}
\caption{\label{fig:helena.and.localeq}
Pressure $p$, normalized surface-average toroidal current density $\langle
J_\text{norm} \rangle$, and safety factor $q$ (a); fitted coefficients $S_0$,
$\sym{\Delta}$, $\sym{\kappa}$, $\asym{\kappa}$, $\sym{\eta}$, and $\asym{\eta}$
(b); numerical magnetic surfaces [(c), solid lines] and analytical ones [(c),
large dots]; vessel outline from reference~\onlinecite{streibl.2003}.}
\end{figure}

\section{Model accuracy and limitations}
\label{sec:accuracy}

Whenever numerical solutions of the GS equation are replaced by analytical
equilibrium models, because the former are not available or its use is not
convenient, then it is necessary to understand the limitations of the latter and
also which equilibrium features are retained in the simplified description. In
this section, the ability of the model~\eqref{eq:solovev.solution.rtheta} to
describe experimentally relevant scenarios is illustrated with a numerical
equilibrium computed by \helena~\cite{huysmans.1991} for parameters typical of
ASDEX-Upgrade operation~\cite{streibl.2003}. The plasma profiles are
\begin{equation}
\frac{dp}{d\Psi} = -1.73 \times 10^6 (1 - \psi)^3,\quad
F\frac{dF}{d\Psi} = 2.13 \, (1 - 4 \psi) (1 - \psi),
\label{eq:equilibrium_profiles}
\end{equation}
both in SI units, and the boundary shape is devised in order to fit the vessel.
Other parameters are the total current $I_\text{p} = 1 \, \text{MA}$,
$F_\text{vac} = 3.3 \, \text{Tm}$, and $\varepsilon = 0.3$. The magnetic axis is
at $R_0 = 1.7 \, \text{m}$, where $B_0 = 1.96 \, \text{T}$.

The equilibrium pressure and toroidal current-density profiles are plotted in
figure~\ref{fig:helena.and.localeq} in terms of the radial-like variable defined
as $\rho_\text{pol} = \sqrt{\psi}$, along with a few magnetic surfaces. For each
surface, labelled by $\psi_i$ ($1 \leqslant i \leqslant 20$), the set of pairs
$r_{ij}$ and $\theta_{ij}$ ($1 \leqslant j \leqslant 200$) returned by \helena\
such that $\psi(r_{ij}, \theta_{ij}) = \psi_i$ is used to retrieve the geometric
coefficients in the model~\eqref{eq:solovev.solution.rtheta} by a least-square
fitting procedure. The fitted coefficients display a mild radial variation,
which validates the local approach. Also, the magnetic surfaces predicted by the
analytical parametrisation~\eqref{eq:magnetic.surface} are seen, again in
figure~\ref{fig:helena.and.localeq}, to be in good agreement with the numerical
ones, showing that the latter's geometry has been suitably captured. However,
such agreement is expected to degrade as one gets closer to the separatrix, as
hinted by the slight mismatch in the outermost surface caused by the limited
number of harmonics ($4$ even and $4$ odd) available in
equations~\eqref{eq:capital.thetas}. More homogeneous terms in
series~\eqref{eq:homogeneous.sum} lead to extra harmonics, but the enhanced
accuracy is outweighed by the increasing complexity in analytical expressions.

\begin{figure}
\includegraphics[scale=1.0]{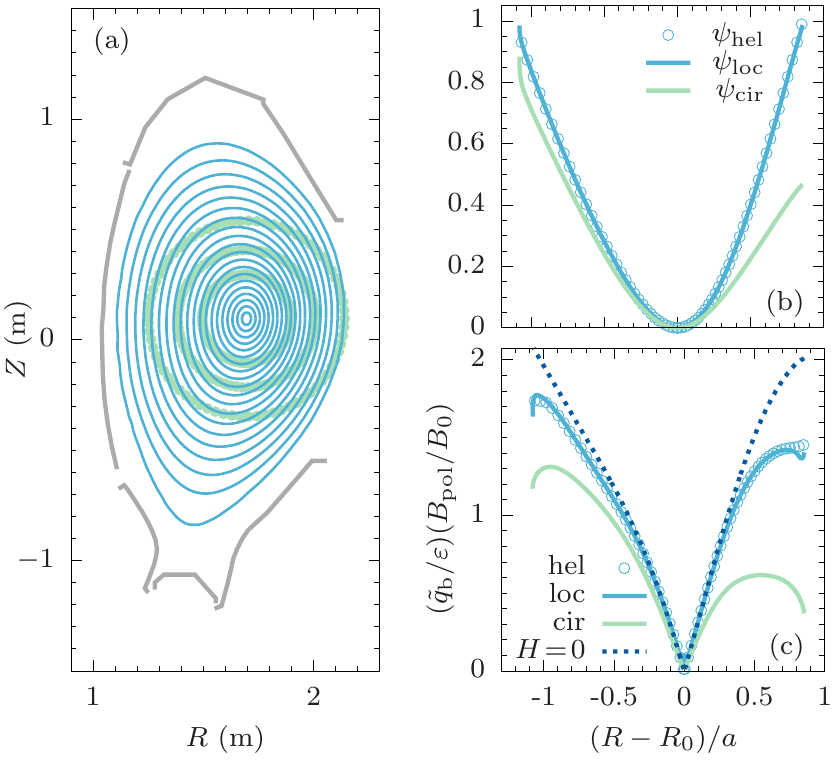}
\caption{\label{fig:helena.and.circular}
Magnetic surfaces from the numerical equilibrium [(a), solid lines] and from the
circular model [(a), large dots]; Normalised poloidal-flux (b) and
poloidal-field magnitude (c) along the midplane as predicted by the circular
model, the local model (with and without the implicit dependence conveyed by
$H$), and as computed by \helena.}
\end{figure}

An equilibrium model widely adopted for analytical work has toroidal magnetic
surfaces with circular and concentric section, for which the field
components are~\cite{lapillonne.2009}
\begin{equation}
B^r = 0, \quad B^\theta = \frac{B_0}{\qb r R} \frac{d\psi}{dr}, \quad
  B_\phi = R_0 B_0,
\label{eq:circular.fields}
\end{equation}
the radial poloidal flux follows the differential equation
\begin{equation}
\frac{d\psi}{dr} = \frac{\qb}{q(r)} \frac{r}{\sqrt{1 - \varepsilon^2 r^2}},
\label{eq:circular.flux}
\end{equation}
and $q(r)$ is the safety factor. Its limitations are evident in
figure~\ref{fig:helena.and.circular}, where the circular surfaces are seen to
depart considerably from the numerical ones. Matching $q(r)$ to the safety
factor computed by \helena\ along the low-field side of the midplane allows
equation~\eqref{eq:circular.flux} to be solved for the poloidal flux, which is
also plotted and seen to deviate from the numerical results. In stark contrast,
the predictions from the local model~\eqref{eq:solovev.solution.rtheta} closely
follow \helena's output regarding the poloidal flux and the poloidal-field
magnitude defined as $B_\text{pol}^2 = B_r B^r + B_\theta B^\theta$.  Models
with locally adjustable coefficients are more flexible to capture local
equilibrium features than global solutions like the circular model, as
figure~\ref{fig:helena.and.circular} illustrates. Yet, such ability requires
geometric-coefficient variations across magnetic surfaces to be taken into
account, as proposed in earlier models~\cite{miller.1998,yu.2012}. Here, this
contribution is accounted for by the factor $H$ in
definition~\eqref{eq:implicit.derivative} and the needed derivatives are
evaluated as finite differences between coefficient values fitted on adjacent
surfaces.

\begin{figure}
\includegraphics[scale=1.0]{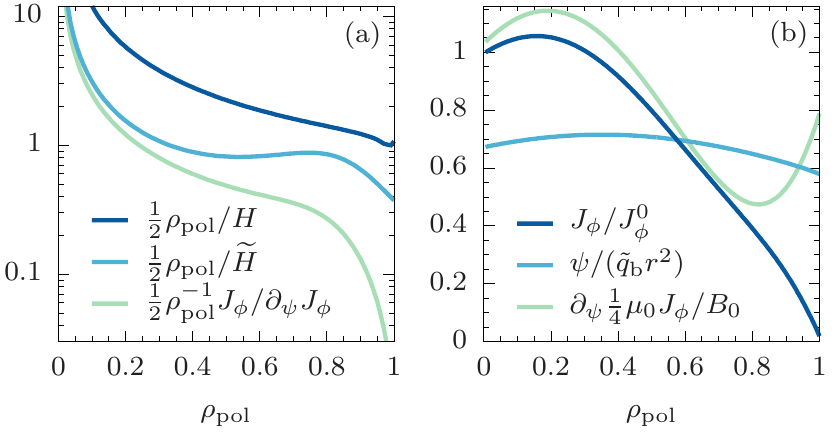}
\caption{\label{fig:validity.size} Asymptotic limits for the size $\bigl| \Delta
\rhopol \bigr|$ of the local model validity domain (a) and radial profile of the
covariant toroidal current density and its partial derivative (b).}
\end{figure}

The magnitude of $H$ also places a limit on the size $\bigl|\Delta\psi\bigr|$ of
the region around a surface $\psi_i$ where the
model~\eqref{eq:solovev.solution.rtheta} with \emph{constant} coefficients is
valid, the condition being
\begin{equation}
\bigl|H \psi_i^{-1} \, \Delta\psi \bigr| =
  \bigl|2H \rhopol^{-1} \, \Delta\rhopol \bigr| \ll 1.
\label{eq:H.condition}
\end{equation}
The value $H$ can be related with the derivative of the toroidal current density
by expanding the GS equation, $J_\phi(R,\psi)$, and each geometric coefficient
in equation~\eqref{eq:solovev.solution.rtheta} around $\psi_i$. The terms linear
in $\Delta\psi$ yield the relation
\begin{equation}
-\Delta \psi \: \partial_\Psi \mu_0 J_\phi\bigl(R, \psi_i \bigr) =
  R \nabla \cdot \bigl( R^{-1} \nabla H \Delta\psi \bigr) \sim
    \frac{4 \widetilde{H} \Delta\psi}{a^2 r^2},
\label{eq:nablaH}
\end{equation}
whence the lowest-order estimate
\begin{equation}
\widetilde{H} \sim
  \qb r^2 \: \partial_\psi \frac{\mu_0 J_\phi(R, \psi_i)}{4 B_0}.
\label{eq:estimateHtilde}
\end{equation}
Figure~\ref{fig:validity.size} displays the asymptotic limits of the size
$\bigl| \Delta \rhopol \bigr|$ set by conditions~\eqref{eq:current.condition}
and~\eqref{eq:H.condition}, the latter using $H$ from
equation~\eqref{eq:implicit.derivative} and, alternatively, $\widetilde{H}$ from
estimate~\eqref{eq:estimateHtilde}. The former condition produces very small
values near the edge ($\rhopol \gtrsim 0.8$) because it is proportional to
$J_\phi$, which approaches zero there. Conversely, the condition that depends on
$\widetilde{H}$, and thus on the dimensionless derivative $\partial_\psi
\tfrac{1}{4} B_0^{-1} \mu_0 J_\phi$ is more robust. Still, both predict smaller
sizes than those found with a numerically evaluated $H$. The reason lies in the
homogeneous solutions that are kept in the
model~\eqref{eq:solovev.solution.rtheta} and its derivatives, but not in
equations~\eqref{eq:current.condition} and~\eqref{eq:estimateHtilde} because $R
\nabla \cdot \bigl(R^{-1} \nabla \psi_\text{h} \bigr) = 0$. The partial
derivative of the toroidal current-density and the ratio $\psi/(\qb r^2)$, both
plotted in figure~\ref{fig:validity.size}, have the same order for the
equilibrium considered here and keep $\widetilde{H} \sim \psi$, which may not be
true in more general cases. Equilibria with larger $J_\phi$ variations require a
smaller $\bigl|\Delta \rhopol \bigr|$ around such locations, but this does not
prevent the local model~\eqref{eq:solovev.solution.rtheta} to apply elsewhere
over the plasma cross section with more favourable validity domains.

\section{Analytical applications}
\label{sec:applications}

Straight-field coordinates $\bigl( \psi, \chi, \phi \bigr)$, where the poloidal
angle $\chi(r, \theta)$ is defined such that $\mathbf{b} = \mathbf{B} / B$
follows
\begin{equation}
b^\phi = q(\psi) b^\chi,
\label{eq:straight.field.condition}
\end{equation}
are a key element in many MHD stability
codes~\cite{mikhailovskii.1997,kerner.1998,mikhailovskii.1998}. Usually,
$\chi(r,\theta)$ is computed from numeric equilibria, but its analytical
evaluation brings insight on how geometric coefficients affect $\kpar =
\mathbf{k} \cdot \mathbf{b}$ and $\kperp^2 = k^2 - \kpar^2$ of a MHD
perturbation with $k_\chi = m$ and $k_\phi = n$. For simplicity, $H =0$ is
assumed henceforth. Finite magnetic-shear effects are kept by expanding
$q(\psi)$ around $\psi_i$ as
\begin{equation}
q(\psi) = q_i + q_i' \bigl( \psi - \psi_i \bigr) + \cdots
    = \frac{\qb}{2 S_0} \biggl(
        \frac{1}{\miota} + \xi \tilde{s}^2 + \cdots \biggr),
\label{eq:q.expansion}
\end{equation}
with $1 / \miota = 2 S_0 \bigl( q_i / \qb \bigr) \bigl( 1 - \mshear \bigr)$,
$\xi = 2 \bigl( S_0 / \psi_i \bigr) \bigl( q_i / \qb \bigr) \mshear$, and
$\mshear = \psi_i q_i'/q_i $, while the solution is sought as a power series in
the small parameters $\varepsilon$ and $\xi$,
\begin{equation}
\chi(r,\theta) = \chi_0(\theta) +
    \xi r^2 \: \chi_\xi(\theta) +
        \varepsilon r \: \chi_\varepsilon(\theta) + \cdots.
\label{eq:ansatz.chi}
\end{equation}
Replacing $b^\chi = b^r \partial_r \chi + b^\theta \partial_\theta \chi$ and the
fields~\eqref{eq:contravariant.Br},~\eqref{eq:contravariant.Btheta},
and~\eqref{eq:covariant.Btor} in definition~\eqref{eq:straight.field.condition}
produces, after collecting the same powers of $\varepsilon$ and $\xi$, the
coupled differential system
\begin{equation}
\begin{gathered}
\Theta_0 \chi_0' = \miota, \\
\Theta_0 \chi_\xi' - \Theta_0' \chi_\xi = - \miota \Theta_0^2 \chi_0', \\
\Theta_0 \chi_\varepsilon'
    - \tfrac{1}{2} \Theta_0' \chi_\varepsilon
      = \bigl(
          \Theta_0 \cos \theta - \tfrac{3}{2} \Theta_1 \bigr) \chi_0'
      - 2 \miota \cos \theta.
\end{gathered}
\label{eq:chi.ode.set}
\end{equation}
Setting the condition $\chi(r,0) = 0$, the solutions are
\begin{equation}
\begin{gathered}
\chi_\xi = - \miota \chi_0 \Theta_0,\\
\chi_0 = \frac{\miota}{\tilde{\kappa}} \biggl[ \arctan
    \frac{\asym{\kappa} + \bigl( 1 - \sym{\kappa} \bigr) \tan
    \theta}{\tilde{\kappa}}
        - \arctan \frac{\asym{\kappa}}{\tilde{\kappa}} \biggr],\\
\chi_\varepsilon = \miota \frac{
    C_0
    + C_1 \cos \theta
    + S_1 \sin \theta
    + C_3 \cos 3 \theta
    + S_3 \sin 3 \theta}{8 \tilde{\kappa}^4 \Theta_0},
\end{gathered}
\label{eq:chi.solutions}
\end{equation}
where $\tilde{\kappa}$ is such that $\tilde{\kappa}^2 + \sym{\kappa}^2 +
\asym{\kappa}^2 = 1$, while $\Theta_{00} = \Theta_0(0)$, $\Theta_{0
\tfrac{\pi}{2}} = \Theta_0(\tfrac{\pi}{2})$, $\Theta_{10} = \Theta_1(0)$, and
also
\begin{widetext}
\begin{equation}
\begin{aligned}
C_0 &=
    4 \bigl( 5 \sym{\kappa} - 1 \bigr) \Theta_{00}^2 \asym{\eta}
    + 9 \bar{\kappa}^2 \Theta_{00} \asym{\kappa}
    - 4 \Theta_{00} \bigl[ 1 - 5 \sym{\Delta} + 3 \sym{\eta} + \sym{\kappa}
    \bigl( 1 + \sym{\Delta} + 9 \sym{\eta} \bigr) \bigr] \asym{\kappa}
    - 12 \Theta_{00} \asym{\eta} \asym{\kappa}^2 - 4 \Theta_{10}
    \asym{\kappa}^3, \\
\tfrac{1}{3} C_1 &=
    -4 \Theta^2_{00} \sym{\kappa} \asym{\eta}
    - \bigl[ 3 + 8 \sym{\Delta} - 4 \sym{\eta} - \sym{\kappa} \bigl(
        2 + 5 \sym{\kappa} \bigr) - 4 \bigl( 4 + \sym{\kappa} \bigr)
        \sym{\kappa} \sym{\eta} \bigr] \asym{\kappa}
    + 4 \bigl( 2 - \sym{\kappa} \bigr) \asym{\eta} \asym{\kappa}^2
    + \bigl( 5 + 4 \sym{\eta} \bigr) \asym{\kappa}^3,\\
\tfrac{1}{2} S_1 &=
    - 2 \Theta_{0\tfrac{\pi}{2}}^2 \bigl[
        2 + \Theta_{0\tfrac{\pi}{2}} \sym{\kappa} + 3 \bigl(
            \sym{\Delta} - \sym{\kappa} \sym{\eta} \bigr) \bigr]
    + 6 \bigl[ 1 - \sym{\kappa} \bigl( 4 - \sym{\kappa} \bigr)
        \bigr] \asym{\eta} \asym{\kappa}
    + \bigl[ 5 - 6 \bigl( \sym{\kappa} + \sym{\Delta} \bigr)
        + 6 \sym{\eta} \bigl( 2 + \sym{\kappa} \bigr)
        + 4 \sym{\kappa}^2 \bigr] \asym{\kappa}^2
    + 6 \asym{\eta} \asym{\kappa}^3 + 2 \asym{\kappa}^4,\\
C_3 &=
    4 \bigl(\Theta_{00}^2 - 3 \asym{\kappa}^2 \bigr) \bigl(
        1 - 2 \sym{\kappa} \bigr) \asym{\eta}
    + 4 \bigl( 1 + \sym{\Delta} \bigr) \asym{\kappa}
    - \bigl(6 - 9 \sym{\kappa} - 4 \sym{\Delta}
        + 8 \sym{\eta} \bigr) \asym{\kappa}^3
    - \bigl[ 7 + 16 \sym{\Delta} + \sym{\kappa} \bigl(
        2 - 9 \sym{\kappa} - 4 \sym{\Delta} - 24 \sym{\eta} \bigr) \bigr]
            \sym{\kappa} \asym{\kappa},\\
S_3 &=
    - 4 \Theta^2_{0\tfrac{\pi}{2}} \bigl[
        \sym{\eta} + \sym{\kappa} \bigl( \Theta_{00} + \sym{\Delta}
            + 2 \sym{\eta} \bigr) \bigr]
    - 24 \sym{\kappa}^2 \asym{\eta} \asym{\kappa}
    - \bigl[ 3 + 8 \sym{\Delta} - 12 \sym{\eta} - \sym{\kappa} \bigl(
        8 + \sym{\kappa} - 4 \sym{\Delta} + 24 \sym{\eta} \bigr) \bigr]
        \asym{\kappa}^2
    +8 \asym{\eta} \asym{\kappa}^3 + 5 \asym{\kappa}^4.
\end{aligned}
\label{eq:chi.varepsilon.consts}
\end{equation}
\end{widetext}
If all geometric coefficients except $S_0$ are set to zero, one finds the
simplified transformation
\begin{equation}
\chi(r,\theta) = \miota \bigl(1 - \miota \: \xi r^2 \bigr) \theta
    - \miota \: \varepsilon r \sin \theta + \cdots
\label{eq:circular.chi}
\end{equation}
that reduces to previous results obtained in the circular limit and without
magnetic shear ($\xi = 0$ and $\miota = 1$)~\cite{lapillonne.2009}.

The lowest order terms of $\kpar = m b^\chi + n b^\phi$ are thus
\begin{equation}
\kpar R_0 = \frac{m + n q}{q_i \bigl( 1 - \mshear \bigr)} \biggl(
    1 - \varepsilon r \cos \theta
        - \xi \miota r^2 \Theta_0 + \cdots \biggr),
\label{eq:kpar}
\end{equation}
whose dependence on $\sym{\kappa}$ and $\asym{\kappa}$ via $\Theta_0$ is rather
weak for low magnetic shear ($\xi \sim \mshear \ll 1$). The expression for
$\kperp$ is too complex in practice, but its linearisation around the limit of
very small $\sym{\kappa}$, $\asym{\kappa}$, $\sym{\eta}$, and $\asym{\eta}$
yields
\begin{multline}
\kperp \frac{a r}{\miota m} = 
    1 - \sym{\kappa} \cos 2 \theta - \asym{\kappa} \sin 2 \theta
    - \xi \miota r^2 \bigl( 1 + \theta \dot{\Theta}_0 \bigr) \\
    + \tfrac{3}{4} \varepsilon r \sym{\Delta} \Bigl[
        \sym{\kappa} \bigl( \cos 3 \theta + \cos \theta \bigr)
        + \asym{\kappa} \bigl( \sin 3 \theta + \sin \theta \bigr) \Bigr] \\
    - \varepsilon r \Bigl[
        \bigl( 1 - \sym{\kappa} \bigr) \cos \theta
            + \tfrac{3}{2} \Theta_1 - \asym{\kappa} \sin \theta \Bigr]
    + \cdots.
\label{eq:kperp}
\end{multline}
Unlike $\kpar$, $\kperp$ depends strongly on $\sym{\kappa}$ and $\asym{\kappa}$,
even if $\xi \ll 1$. First-order terms in $\varepsilon$ enhance these
dependencies, couple them with $\sym{\Delta}$, and connect also with
$\sym{\eta}$ and $\asym{\eta}$ via $\Theta_1$.

\section{Conclusions}

In summary, a local magnetic-equilibrium model with up-down asymmetric cross
section was developed, where the poloidal-field flux is expanded as a series of
Solovev solutions with radially changing coefficients. The model is accurate to
fourth-order terms in the inverse aspect ratio and depends on eight free
parameters, one for each independent poloidal-angle harmonic (five even and
three odd), of which three were shown to relate with the conventional
definitions of Shafranov shift, elongation, and triangularity.

In contrast with other local equilibrium models, the proposed approach was
devised to produce analytically tractable expressions for the magnetic-field
components. Despite such requirement, the corresponding magnetic-surface
parametrisation was seen to describe equilibrium shapes, poloidal flux
distributions, and magnetic-field configurations typically found in tokamak
experiments. A size estimate of the domain where a local solution with constant
geometric coefficients is valid was provided in terms of the local toroidal
current-density derivative.

As an example of analytical application, the transformation to straight-field
coordinates was obtained, up to first-order terms in the inverse aspect ratio
and in the normalised magnetic shear, and then used to understand how the values
$\kpar$ and $\kperp$ of a MHD perturbation depend on equilibrium geometry. The
suitability of the proposed local model to assess equilibrium-shaping effects,
as illustrated in the examples provided, is expected to afford useful analytical
insight into a wide variety of tokamak-plasma phenomena.

\section*{Acknowledgments}
\noindent IPFN activities were financially supported by ``Funda\c{c}\~{a}o para
a Ci\^{e}ncia e Tecnologia'' (FCT) through project UID/FIS/50010/2013.

%
\end{document}